\documentstyle[11pt,epsfig]{article}
\begin{document}
\title{\bf Interacting Induced Dark Energy Model}
\author{Amir F.~Bahrehbakhsh\thanks{email: amirfarshad@gmail.com}\\
  {\small \emph{Department of Physics, Faculty of Science, Payam-e-Noor University, Iran}}\\
  {\small \emph{Department of Physics and Astronomy, University of California, Irvine, CA 92697, USA}}}
\date{\small June 25, 2017}
\maketitle
\begin{abstract}
Following the idea of the induced matter theory, for a non--vacuum
five--dimensional version of general relativity, we propose a model
in which the induced terms emerging from the extra dimension in our
four--dimensional space--time, supposed to be as dark energy. Then
we investigate the FLRW type cosmological equations and illustrate
that when the scale factor of the fifth dimension has no dynamics,
in early time the universe expands with deceleration and then in
late time, expands with acceleration. In this case, the state
parameter of the effective dark energy has a range of $-1<\bar{w}_{_{X}}<0$
and it has the value $-1/2$ for present time. The results for current
acceleration impose that $\Omega_{_{X\circ}}>2\Omega_{_{M\circ}}$ which
is in agreement with the measurements. We show that the effective energy
density of dark energy have been having the same order of magnitude of
the effective energy density of matter from the early time in the
decelerating epoch of the universe expansion until now. The model
avoids the cosmological coincidence problem.
\end{abstract}
%
%\medskip
%{\small \noindent
% PACS number: $04.50.-h$\ ; $04.50.Kd$\ ; $04.20.Cv$\ ; $04.90.+e$\ ; $98.80.Jk$}\newline
%
\bigskip
\section{Introduction}
\indent

According to the observations of distant type--Ia supernovae which indicate that
the expansion of the universe is presently accelerating~\cite{Perlmutter:1997zf}--
\cite{Carroll:2003qq}, the universe should mainly be filled with what is usually
called dark energy~\cite{Sahni:2004ai},~\cite{Kamionkowski:2007wv}. Therefore,
a considerable amount of researches has been performed to explain this acceleration
which, in general, can be classified into two groups. The first group contains all
modifications to the standard model of cosmology: cosmological constant
($\Lambda \rm CDM$), modified matter models such as the quintessence~\cite{Zlatev:1998tr}--
\cite{Blais:2004vt}, the k-essence~\cite{ArmendarizPicon:2000ah}--~\cite{Scherrer:2004au},
Chaplygin gas~\cite{Bento:2002ps} models and also the models based on the modified theories
of gravity like $f(R)$ gravity~\cite{Nojiri:2006ri},~\cite{Schmidt:2006jt}, $f(T)$ gravity
~\cite{Li:2010cg},~\cite{Myrzakulov:2010vz} and $f(R,T)$ gravity~\cite{Harko:2011kv}.

The second group includes all the models based on the alternative theories
of gravity such as Brans--Dicke (BD) theories~\cite{Jordan:1955}--~\cite{Fujii:2004},
Kaluza--Klein (KK) theories~\cite{Kaluza:1921}--~\cite{Overduin:1998pn}, induced matter
(IM) theories~\cite{Wesson:1999}--~\cite{Bahrehbakhsh:2017poi}, brane world
scenarios~\cite{Randall:1999ee}--~\cite{Maartens:2010ar}, \"{U}ber gravity~\cite{Khosravi:2017aqq}
and some mixture of these alternative theories~\cite{deLeon:2009nd}--~\cite{Bahrehbakhsh:2013qda}
in addition to many other ones~\cite{Clifton:2011jh},~\cite{Yoo:2012}. Hence, in the recent two
decades, explaining the accelerated expansion of the universe through the fundamental theories
has been a great challenge.

Since, the most successful unifying theories such as string theory and M--theory work in higher
dimensional space--time~\cite{Pavsic:2001qh}, in this work similar to the idea of IM theory,
we try to find a geometrical origin, based on the fifth dimension, for dark energy.
The significant of the IM theories is that inducing $5D$ field equations without matter sources leads
to the $4D$ field equations with matter sources. In another word, the matter sources of $4D$ space--times
can be viewed as a manifestation of extra dimensions~\cite{Wesson:1999}--~\cite{Wesson:2014raa}.

But, here we consider a non--vacuum $5D$
version of general relativity (GR) and propose a model in which, neither the baryonic nor dark matter,
but just only dark energy is considered as an effect of existence of the fifth
dimension. We employ a generalized $5D$ Friedmann--Lema\^{\i}tre--Robertson--Walker
(FLRW) type metric for the model and investigate its cosmological implications.
In section~$2$, we give a brief review of the $5D$ version of GR and induce the
non--vacuum $5D$ field equations in a $4D$ hypersurface and then collect the extra
terms emerging from the fifth dimension as dark energy component of the induced
energy--momentum tensor in addition to the common energy--momentum tensor of the matter
(the baryonic and dark matter). In section~$3$, we derive the FLRW cosmological
equations by considering a generalized FLRW metric in a $5D$ space--time. Then, we obtain
the total energy conservation equation and separate it into two interacting energy
conservation equations one for all kind of the matter and the other one for dark energy.
In the following, we investigate three simple and important cases. First, we assume the
fifth--dimension momentum density, $Q$, equals to zero and then we suppose the pressure
in fifth dimension, $p_{_{5}}$, be zero. Finally, we assume the scale factor of the
fifth dimension to be a constant. At the end, in the last section conclusion is presented.
%%%%%%%%%%%%%%%%%%%%%%%%%%%%%%%%%%%%%%%%%%%%%%%%
\section{Induced Dark Energy Model}
\indent
Following the approach of IM theories, one can consider a $5D$ version of GR with action,
\begin{equation}\label{5D Action}
\emph{S}
=\int\sqrt{|{}^{_{(5)}}g|} \left (\frac{1}{16\pi G}\
^{^{(5)}}\!R+
L_{m} \right )d^{5}x\, ,
\end{equation}
where $c=1$, $^{^{(5)}}R$ is $5D$ Ricci scalar, $^{_{(5)}}g$ is the determinant of $5D$
metric $g_{_{AB}}$ and $\emph{L}_{m}$ represents the matter Lagrangian. Hence, the
Einstein field equations in five dimension can be written as
\begin{equation}\label{5D Einstein Eq}
^{^{(5)}}G_{_{AB}}=8\pi G \
^{^{(5)}}T_{_{AB}} \, ,
\end{equation}
where the capital Latin indices run from zero to four, $^{^{(5)}}G_{_{AB}}$ is $5D$
Einstein tensor and $^{^{(5)}}T_{_{AB}}$ is $5D$ energy--momentum tensor.
As a reasonable assumption we consider that $^{^{(5)}}T_{\alpha\beta}$
(the Greek indices go from zero to three) represents the same baryonic and
dark matter sources of a $4D$ hypersurface, i.e. $T^{^{(M)}}_{\alpha\beta}$.
We also take the $5D$ energy--momentum tensor in the form of
\begin{equation}\label{T5}
^{^{(5)}}T_{_{AB}}=\left( \begin {array}{ccccc}\rho_{_{M}}&0&0&0&Q
\\0&\ p_{_{M}}&0&0&0
\\0&0& p_{_{M}}&0&0
\\0&0&0&\ p_{_{M}}&0
\\Q&0&0&0& p_{_{5}}
\end {array} \right),
\end{equation}
where $\rho_{_{M}}$ and $p_{_{M}}$ are energy density and pressure of the matter
(baryonic and dark matter) and also $p_{_{5}}$ and $Q$ respectively represent
the pressure and momentum density in fifth dimension. For cosmological purposes
we restricts our attention to the $5D$ warped metrics of the form
\begin{eqnarray}\label{5D Metric}
dS^2=g_{_{AB}}(x^{C})dx^{A}dx^{B}=^{^{(5)}}\!g_{\mu\nu}(x^{C})dx^{\mu}dx^{\nu}+g_{_{44}}(x^{C})dy^2
\equiv ^{^{(5)}}\!g_{\mu\nu}(x^{C})dx^{\mu}dx^{\nu}+\epsilon
b^{2}(x^{C})dy^2\, ,
\end{eqnarray}
in local coordinates $x^{A}=(x^{\mu},y)$,
where $y$ represents the fifth coordinate and $\epsilon^2=1$.
By assuming the $5D$ space--time is foliated
by a family of hypersurfaces, $\Sigma$, which are defined by fixed values of
$y$, then, one can obtain the intrinsic metric of any typical
hypersurface, e.g. $\Sigma_{\circ}(y=y_{\circ})$, by
restricting the line element (\ref{5D Metric}) to displacements confined
to it. Therefore, the induced metric on the hypersurface $\Sigma_{\circ}$
becomes
\begin{eqnarray}\label{4D Metric}
ds^2=\
^{^{(5)}}\!g_{\mu\nu}(x^{\alpha},y_{\circ})dx^{\mu}dx^{\nu}\equiv
g_{\mu\nu}dx^{\mu}dx^{\nu}\, ,
\end{eqnarray}
thus the usual $4D$ space--time metric, $g_{\mu\nu}$, can be recovered.
Therefore after some manipulations, equation (\ref{5D Einstein Eq}) on the hypersurface $\Sigma_{\circ}$
can be written as
\begin{eqnarray}\label{4D Einstein Eq}
G_{\alpha\beta}=8\pi G(T^{^{(M)}}_{\alpha\beta}+T^{^{(X)}}_{\alpha\beta}),
\end{eqnarray}
where we consider $T^{^{(X)}}_{\alpha\beta}$ as dark energy
component of the energy--momentum tensor that analogous to the IM theory
\cite{Bahrehbakhsh:2010cx} and \cite{Bahrehbakhsh:2013qda}, defined by
\begin{eqnarray}\label{T X}
T^{^{(X)}}_{\alpha\beta}\equiv T^{^{\rm(IM)}}_{\alpha\beta}
\equiv\frac{1}{8\pi G}\Bigg \{  \frac{b_{;\alpha\beta}}{b}-\frac{\Box b}{b}g_{\alpha\beta}
-\frac{\epsilon}{2b^2}\Bigg [\frac{b'}{b}g'_{\alpha\beta} -g''_{\alpha\beta}
+g^{\mu\nu}g'_{\alpha\mu}g'_{\beta\nu}-
\frac{1}{2}g^{\mu\nu}g'_{\mu\nu}g'_{\alpha\beta}\\ \nonumber-g_{\alpha\beta}\bigg (\frac{b'}{b}g^{\mu\nu}g'_{\mu\nu}
-g^{\mu\nu}g''_{\mu\nu}-\frac{1}{4}g^{\mu\nu}g^{\rho\sigma}g'_{\mu\nu}g'_{\rho\sigma}
-\frac{3}{4}g'^{\mu\nu}g'_{\mu\nu} \bigg )\Bigg ]
\Bigg \}\, ,
\end{eqnarray}
in which the prime denotes derivative with respect to the fifth coordinate.

In the following section, we consider a generalized FLRW metric in
a $5D$ universe and investigate its cosmological properties.
%%%%%%%%%%%%%%%%%%%%%%%%%%%
\section{Generalized FLRW Cosmology}
\indent

For a $5D$ universe with an extra
space--like dimension in addition to the three usual spatially homogenous
and isotropic ones, metric (\ref{5D Metric}), as a generalized FLRW solution,
can be written as
\begin{eqnarray}\label{FLRW Metric1}
dS^2=-dt^2+\tilde{a}^2(t,\tilde{y})\left [\frac{dr^2}{1-kr^2}+r^2(d\theta^2+\sin^2\theta
d\varphi^2)\right ]+\tilde{b}^2(t,\tilde{y})d\tilde{y}^2\, .
\end{eqnarray}
Generally, the scale factors $\tilde{a}$ and $\tilde{b}$ should be functions of
cosmic time and extra dimension coordinate. However, for physical plausibility
and simplicity, we assume that they are separable functions of time and extra coordinate.
Besides, the functionality of the scale factor $\tilde{b}$ on $\tilde{y}$ could be eliminated
by transforming to a new extra coordinate $y$, hence metric (\ref{FLRW Metric1}) can be rewritten as
\begin{eqnarray}\label{FLRW Metric2}
dS^2=-dt^2+ a^2(t)l^2(y)\left [\frac{dr^2}{1-kr^2}+r^2(d\theta^2+\sin^2\theta
d\varphi^2)\right ]+b^2(t)dy^2\, .
\end{eqnarray}
By considering metric (\ref{FLRW Metric2}) and assuming $H\equiv\dot{a}/a$,
$B\equiv\dot{b}/b$ and $L\equiv{l'}/l$, the Einstein equations
(\ref{5D Einstein Eq}) reduce as follows. The time component, $A=0=B$, provides
\begin{eqnarray}\label{FLRW1}
H^2=\frac{8\pi G}{3}\rho_{_{M}}-HB+\frac{1}{b^2}(L'+2L^2)-\frac{k}{a^2 l^2}
\equiv\frac{8\pi G}{3}\tilde{\rho}-\frac{k}{a^2 l^2}\, ,
\end{eqnarray}
the spatial ones, $A=B=1,2 \ \rm or \ 3$, give
\begin{eqnarray}\label{FLRW2}
\frac{\ddot{a}}{a}=-4\pi Gp_{_{M}}-\frac{1}{2}H^{2}
-HB-\frac{1}{2}(\dot{B}+B^{2})-\frac{1}{2b^2}(2L'+3L^2)-\frac{k}{2a^2l^2}
\equiv-\frac{4\pi G}{3} (\tilde{\rho}+3\tilde{p})\, ,
\end{eqnarray}
the $A=4=B$ component yields
\begin{eqnarray}\label{FLRW3}
\frac{\ddot{a}}{a}=\frac{8\pi G}{3}p_{_{5}}-H^{2}+\frac{1}{b^2}L^{2}-\frac{k}{a^2l^2}
\end{eqnarray}
and finally $A=0$ and $B=4$ (and also $A=4$ and $B=0$) component results
\begin{eqnarray}\label{FLRW4}
L(H-B)=\frac{8\pi G}{3}Q\, .
\end{eqnarray}
In equations (\ref{FLRW1}) and (\ref{FLRW2}) we have defined
$\tilde{\rho}\equiv\rho_{_{M}}+\rho_{_{X}}$ and $\tilde{p}\equiv p_{_{M}}+p_{_{X}}$
in which the energy density and pressure of dark energy has specified by
employing relation (\ref{4D Einstein Eq}), as
\begin{eqnarray}\label{Ro X}
\rho_{_{X}}\equiv T^{^{(X)}}_{_{\circ\circ}}=\frac{3}{8\pi G}\Big [\frac{1}{b^2}(L'+2L^2)-HB \Big ]
\end{eqnarray}
and
\begin{eqnarray}\label{P X}
p_{_{X}}\equiv -T^{^{(X)}}_{_{ii}}=\frac{1}{8\pi G}\Big[\dot{B}+B^2+2HB-\frac{1}{b^2}(2L'+3L^2) \Big].
\end{eqnarray}

By substituting equation (\ref{FLRW1}) in (\ref{FLRW3}) and assuming $p_{_{M}}=0$
for current important components of the matter (baryonic and dark matter), one has
\begin{eqnarray}\label{NFLRW3 Result}
b^2 \Big[\dot{H}+H^2-HB+\frac{8\pi G}{3}(\rho_{_{M}}-p_{_{5}})\Big]=-(L'+L^{2})=-\lambda^2\, ,
\end{eqnarray}
where $\lambda$ is a non--zero real constant and obtained from the fact that the left hand side
of the equation is a function of cosmic time but the right hand side of the equation is a function
of extra dimension coordinate, thus both sides of the equation must be equal to a constant.
From the equation (\ref{NFLRW3 Result}) one can get
\begin{eqnarray}\label{L}
l(y)=l_{\circ}e^{\pm \lambda(y-y_{_{\circ}})}\, ,
\end{eqnarray}
with $L=\pm\lambda$ and $L'=0$.
%As we have supposed that $l(y)$ and thereupon
%$L$ are real functions, hence $C$ is positive real constants.
Hence, the equations (\ref{FLRW1})--(\ref{P X}) on the hypersurface
$\Sigma_{\circ}(y=y_{_{\circ}})$ can be rewritten as
\begin{eqnarray}\label{NFLRW1}
H^2=\frac{8\pi G}{3}\rho_{_{M}}-HB+\frac{2\lambda^2}{b^2}-\frac{k}{a^2 l_{\circ}^2}
\equiv\frac{8\pi G}{3}\tilde{\rho}-\frac{k}{a^2 l_{\circ}^2}\, ,
\end{eqnarray}
\begin{eqnarray}\label{NFLRW3}
\frac{\ddot{a}}{a}=\frac{8\pi G}{3}p_{_{5}}-H^{2}+\frac{\lambda^2}{b^2}-\frac{k}{a^2l_{\circ}^2}
\equiv-\frac{4\pi G}{3} (\tilde{\rho}+3\tilde{p})\, ,
\end{eqnarray}
\begin{eqnarray}\label{NFLRW4}
H-B=\frac{8\pi G}{3\lambda}Q\, ,
\end{eqnarray}
\begin{eqnarray}\label{NRo X}
\rho_{_{X}}=\frac{3}{8\pi G}\Big (\frac{2\lambda^2}{b^2}-HB \Big )
\end{eqnarray}
and
\begin{eqnarray}\label{NP X}
p_{_{X}}=\frac{1}{8\pi G}\Big(\dot{B}+B^2+2HB-\frac{3\lambda^2}{b^2}\Big)\, ,
\end{eqnarray}
in which
\begin{eqnarray}\label{P5}
p_{_{5}}=\frac{1}{2}(\rho_{_{M}}+\rho_{_{X}}-3p_{_{X}})-\frac{3\lambda^2}{8\pi G b^2}\, .
\end{eqnarray}
On the other hand, to obtain energy conservation equation,
one can take the time derivative of the equation (\ref{FLRW1})
and substitute the equation (\ref{FLRW2}) into it and get
\begin{eqnarray}\label{Conservation MIX}
\dot{\tilde{\rho}}+3H(\tilde{\rho}+\tilde{p})=0\, .
\end{eqnarray}
As the dark energy density is of the same order as the dark matter energy density
in the present universe, one can imagine that there is some connection between
dark energy and dark matter and expect their conservation equations may not to
be independent. Hence, we assume equation (\ref{Conservation MIX}) can plausibly
separate into two distinguished equations for $\rho_{_{X}}$ and $\rho_{_{M}}$ as
\begin{eqnarray}\label{Conservation X}
\dot{\rho}_{_{X}}+3H(\rho_{_{X}}+p_{_{X}})=f(t)
\end{eqnarray}
and
\begin{eqnarray}\label{Conservation M}
\dot{\rho}_{_{M}}+3H(\rho_{_{M}}+p_{_{M}})=-f(t)\, ,
\end{eqnarray}
where $f(t)$ is assumed to be the interacting term between dark energy and dark matter.

Now, substituting relations (\ref{NRo X}) and (\ref{NP X}) into the equation
(\ref{Conservation X}), gets
\begin{eqnarray}\label{Conservation X Result}
b^2 \Big[\dot{H}+H^2-HB+\frac{8\pi G}{3}\frac{f(t)}{B}\Big]=\frac{\lambda^2}{B}(3H-4B)\, ,
\end{eqnarray}
which comparing it with the equation (\ref{NFLRW3 Result}) results
\begin{eqnarray}\label{f}
f(t)=B(\rho_{_{M}}-p_{_{5}})+\frac{3\lambda Q}{b^2}\, .
\end{eqnarray}
In the following, we investigate three cases. First, we assume the fifth--dimension
momentum density be zero, i.e. $Q=0$. Then we suppose the pressure in fifth dimension,
 $p_{_{5}}$, equals to zero. Finally, we assume the scale factor of the fifth dimension
 to be a constant.
\\
\subsection{The Model with Zero Fifth--Dimension Momentum Density; $Q=0$}
\indent

By considering $Q=0$, from the equation (\ref{NFLRW4}) one has $B=H$ with
$b=(a_{_{\circ}}/b_{_{\circ}})a$. Hence the equations (\ref{NFLRW1}) and
(\ref{NFLRW3}) and the interacting term between matter and dark energy become
\begin{eqnarray}\label{NNFLRW1}
H^2=\frac{4\pi G}{3}\rho_{_{M}}+\frac{a_{_{\circ}}^2\lambda^2}{b_{_{\circ}}^2a^2}-\frac{k}{2l_{\circ}^2 a^2}\, ,
\end{eqnarray}
\begin{eqnarray}\label{NNFLRW3}
\frac{\ddot{a}}{a}=-\frac{4\pi G}{3}(\rho_{_{M}}-2p_{_{5}})-\frac{k}{2l_{\circ}^2a^2}\,
\end{eqnarray}
and
\begin{eqnarray}\label{f1}
f(t)=H(\rho_{_{M}}-p_{_{5}})\, .
\end{eqnarray}
In the case $f(t)=0$, one has $p_{_{5}}=\rho_{_{M}}$ and from the equations
(\ref{Conservation M}) gets
\begin{eqnarray}\label{Ro M1}
\rho_{_{M}}=\rho_{_{M\circ}} \Big (\frac{a_{\circ}}{a}\Big)^3
\end{eqnarray}
and hence equation (\ref{NNFLRW3}) becomes
\begin{eqnarray}\label{NNFLRW3A}
\frac{\ddot{a}}{a}=\frac{4\pi G\rho_{_{M\circ}}}{3} \Big (\frac{a_{\circ}}{a}\Big)^3 -\frac{k}{2l_{\circ}^2a^2}\, .
\end{eqnarray}
For spatially flat or open universe ($k=0$ and $-1$), equation (\ref{NNFLRW3A})
illustrates eternal accelerated expansion, but for a closed one, it suggests that
the expansion of the universe first accelerates and then decelerates. Therefore,
the non--interacting case for all kind of the geometry has results contrary to the
observations and hence it is not a suitable case.

When $f(t)\neq0$, from the acceleration equation (\ref{NNFLRW3}), it is clear that
the first term containing energy density of matter, $\rho_{_{M}}$, only contributes
in deceleration of the universe expansion. However, the second and third terms are
capable to contributes in the acceleration but, multiplying the equation (\ref{NNFLRW1})
by $a^2$, shows that these terms also can not increase the velocity of the universe
expansion, i.e. $\dot{a}$. Hence the model, when the fifth--dimension momentum density
is zero, is not self consistent to having the accelerating solutions. Therefore, in the
following as the second case we assume the fifth--dimension pressure to be zero.
\\
\subsection{The Model with Zero Fifth--Dimension Pressure; $p_{_{5}}=0$}
\indent

Taking $p_{_{5}}=0$ with a mathematically simplification and physically plausible
assumption as $B=nH$ gives
\begin{eqnarray}\label{Q2}
Q=\frac{3\lambda}{8\pi G}(1-n)H
\end{eqnarray}
and
\begin{eqnarray}\label{f2}
f(t)=\Big(n\rho_{_{M}}+\frac{9\lambda^2}{8\pi G}\frac{1-n}{a^{2n}}\Big)H\, ,
\end{eqnarray}
which substituting the second one in the equation (\ref{Conservation M}) leads
\begin{eqnarray}\label{Ro M2}
\rho_{_{M}}=\rho_{_{M\circ}} \Big (\frac{a_{\circ}}{a}\Big)^{n+3}+\beta\frac{n-1}{3-n}a^{-2n}\, .
\end{eqnarray}
Then, equations (\ref{NFLRW1}) and (\ref{NFLRW3}) become
\begin{eqnarray}\label{NNNFLRW1}
H^2=\frac{8\pi G \rho_{_{M\circ}}}{3(n+1)} \Big (\frac{a_{\circ}}{a}\Big)^{n+3}+
\frac{n+3}{(n+1)(3-n)}\frac{\lambda^2}{a^{2n}}-\frac{k}{(n+1)l_{\circ}^2a^{2}}\,
\end{eqnarray}
and
\begin{eqnarray}\label{NNNFLRW3}
\frac{\ddot{a}}{a}=-\frac{8\pi G \rho_{_{M\circ}}}{3(n+1)} \Big (\frac{a_{\circ}}{a}\Big)^{n+3}+
\frac{n(1-n)}{(n+1)(3-n)}\frac{\lambda^2}{a^{2n}}-\frac{n k}{(n+1)l_{\circ}^2a^{2}}\, .
\end{eqnarray}
The consistency condition between the equations (\ref{f2}) yields to $n=1$, which gives
the same unsuitable results of the previous case with $Q=0$. Therefore, as the last assumption
we consider that the scale factor of the fifth dimension does not have any dynamics.
\\
\subsection{The Model with Static Scale Factor of the Fifth--Dimension; $b=\rm constant$}
\indent

Assuming $b=b_{\circ}=cte$, gets $B=0$ and from the equations (\ref{NFLRW1})--(\ref{NP X}) one has
\begin{eqnarray}\label{NNNNFLRW1}
H^2=\frac{8\pi G}{3}(\rho_{_{M}}+\rho_{_{X}})-\frac{k}{l_{\circ}^2 a^2}\, ,
\end{eqnarray}
\begin{eqnarray}\label{NNNNFLRW3}
\frac{\ddot{a}}{a}=-\frac{4\pi G}{3}(\rho_{_{M}}+\rho_{_{X}}+3p_{_{X}})\,
\end{eqnarray}
and
\begin{eqnarray}\label{Q3}
Q=\frac{3\lambda}{8\pi G}H\, ,
\end{eqnarray}
in which
\begin{eqnarray}\label{Ro X3}
\rho_{_{X}}=\frac{3\lambda^2}{4\pi G b_{\circ}^2}
\end{eqnarray}
and
\begin{eqnarray}\label{P X3}
p_{_{X}}=-\frac{3\lambda^2}{8\pi G b_{\circ}^2}\, .
\end{eqnarray}
Also, from the equations (\ref{f}), (\ref{Q3}) and (\ref{Ro X3}) one gets
\begin{eqnarray}\label{f3}
f(t)=\frac{9\lambda^2}{4\pi G b_{\circ}^2}H=\frac{3}{2}\rho_{_{X}}H\, ,
\end{eqnarray}
thereupon, substituting the relation (\ref{f3}) in the equation (\ref{Conservation M}) gives
\begin{eqnarray}\label{Ro M3}
\rho_{_{M}}=\rho_{_{M\circ}} \Big (\frac{a_{\circ}}{a}\Big)^3+
\frac{3\lambda^2}{8\pi G b_{\circ}^2}\Big[\Big (\frac{a_{\circ}}{a}\Big)^3-1 \Big]\, .
\end{eqnarray}
To have the common form of the evolution of energy density of matter,
we define the effective energy densities of matter and dark energy as follows
\begin{eqnarray}\label{NRo M3}
\bar{\rho}_{_{M}}\equiv\rho_{_{M\circ}} \Big (\frac{a_{\circ}}{a}\Big)^3
\end{eqnarray}
and
\begin{eqnarray}\label{NRo X3}
\bar{\rho}_{_{X}}\equiv\frac{3\lambda^2}{8\pi G b_{\circ}^2}\Big[\Big (\frac{a_{\circ}}{a}\Big)^3+1 \Big]\, .
\end{eqnarray}
Hence, the equations (\ref{NNNNFLRW1}) and (\ref{NNNNFLRW3}) can be rewritten as
\begin{eqnarray}\label{5NFLRW1}
H^2=\frac{8\pi G}{3}(\bar{\rho}_{_{M}}+\bar{\rho}_{_{X}})-\frac{k}{l_{\circ}^2 a^2}
\equiv\frac{8\pi G}{3}\rho_{_{M\circ}} \Big (\frac{a_{\circ}}{a}\Big)^3+
\frac{\lambda^2}{b_{\circ}^2}\Big[1+\Big (\frac{a_{\circ}}{a}\Big)^3 \Big]-\frac{k}{l_{\circ}^2 a^2}\,
\end{eqnarray}
and
\begin{eqnarray}\label{5NFLRW3}
\frac{\ddot{a}}{a}=-\frac{4\pi G}{3}(\bar{\rho}_{_{M}}+\bar{\rho}_{_{X}}+3\bar{p}_{_{X}})
\equiv-\frac{4\pi G}{3}\rho_{_{M\circ}} \Big (\frac{a_{\circ}}{a}\Big)^3+
\frac{\lambda^2}{b_{\circ}^2}\Big[ 1-\frac{1}{2}\Big(\frac{a_{\circ}}{a}\Big)^3\Big]\,
\end{eqnarray}
in which
\begin{eqnarray}\label{NP X3}
\bar{p}_{_{X}}=p_{_{X}}=-\frac{3\lambda^2}{8\pi G b_{\circ}^2}\, .
\end{eqnarray}
Then, the state parameter of effective dark energy becomes
\begin{eqnarray}\label{WX}
\bar{w}_{_{X}}=\frac{\bar{p}_{_{X}}}{\bar{\rho}_{_{X}}}=-\frac{a^{3}}{a^{3}+a_{\circ}^{3}}\, ,
\end{eqnarray}
which has a range of $-1<\bar{w}_{_{X}}<0$. Fig.~$1$ shows the changes of state parameter
of effective dark energy.
%%%%%%%%%%%%%%%%%%%%%%%%%%%%%%%%%%%%%%%%%%%33333
\begin{figure}
\begin{center}
\epsfig{figure=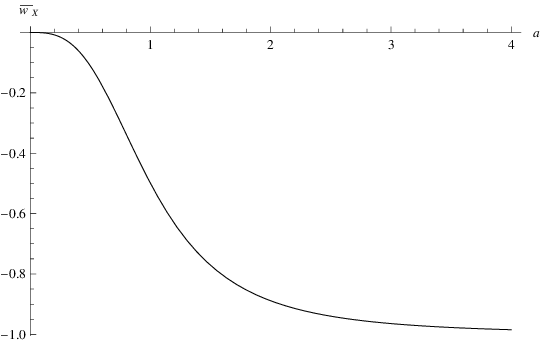} \caption{\footnotesize
Changes of state parameter of effective dark energy with evolution of the scale
factor of the ordinary spatial dimensions by setting $a_{\circ}=1$.}
\end{center}
\end{figure}
%%%%%%%%%%%%%%%%%%%%%%%%%%%%%%%%%%%%%%%%%%%%%%%%%%%%%%%%%%%%%%%%%%%%%%%%%%%%%%%%%%

Since the measurements of anisotropies in the cosmic microwave background indicates that
the universe must be very close to spatially flat one~\cite{Bahcall:1999xn}--\cite{Hanany:2000qf},
then we consider $k=0$. Equation (\ref{5NFLRW3}) illustrates that the universe expansion, in
the early epoch, decelerates and then in the late epoch, accelerates. Present acceleration
imposes $\frac{\lambda^{2}}{b_{\circ}^{2}}>\frac{8\pi G}{3}\rho_{_{M\circ}}$, this yields to
$\Omega_{_{X\circ}}>2\Omega_{_{M\circ}}$, or, $\Omega_{_{X\circ}}>2/3$ and $\Omega_{_{M\circ}}<1/3$,
which is in agreement with the measurements, i.e. $\Omega_{_{X\circ}}\approx0.7$ and $\Omega_{_{M\circ}}\approx0.3$.

Also, one gets
\begin{eqnarray}\label{Rate}
\frac{\bar{\rho}_{_{X}}}{\bar{\rho}_{_{M}}}=\frac{3\lambda^{2}}{8 \pi G b_{\circ}^2\rho_{_{M\circ}} } \frac{1}{1+\bar{w}_{_{X}}}>
\frac{1}{1+\bar{w}_{_{X}}}=\left\{
  \begin{array}{ll}
   1 \quad \ \qquad \bar{w}_{_{X}}\rightarrow0 \qquad \quad \ \ (a\rightarrow0)\\
   2 \quad \ \qquad \bar{w}_{_{X}}=-1/2 \qquad (a=a_{\circ})\\
   +\infty \qquad \bar{w}_{_{X}}\rightarrow-1 \qquad \quad (a\rightarrow \infty)\, ,
  \end{array}
\right.
\end{eqnarray}
which shows that the effective energy density of dark energy dominates all the time of the
universe expansion with the same order of magnitude of energy density of matter in the past and present
time. Hence, the cosmological coincidence problem can be solved with this model.
Note, this domination of dark energy does not mean eternally accelerating expansion in this model,
the term, $-\frac{\lambda^2}{2b_{\circ}^2}(\frac{a_{\circ}}{a})^3$, actually contributes in deceleration, so
at early time beside the energy density of matter, the effective energy density of dark energy
also contributes in deceleration.

At the end, we should mention that the results of GW170817 apply only to large extra dimensions because,
gravitational waves detected from the neutron star collision have wavelengths of thousands of kilometers
and would not affect tiny extra dimensions.
Hence, this imposes that $b_{\circ}$ must be very small.

\section{Conclusions}
\indent

The observations illustrate that the universe is presently in an accelerated
expanding phase. Hence, the main content of the universe should be consisted
of what commonly called dark energy. However, an enormous amount of work has been performed
to explain this acceleration, but the origin and the nature of dark energy is unknown yet.

In this work, following the approach of the induced matter theory, we have investigated
the cosmological implications of a non--vacuum five--dimensional version of general relativity
in order to explain both deceleration and acceleration eras of the universe expansion and solve the
cosmological coincidence problem.
%%%%%%%%%%%%%%%%%%%%%%%%%%%%%%%%%%%%%%%%%%%%%%%%%%%%%%%%%%%%%%%%%%%%%
In this respect, in general we have considered a five dimensional energy--momentum tensor contains
the elements, pressure and momentum density in fifth dimension, i.e., $p_{_{5}}$ and $Q$, in addition
to the energy density and the pressure of the matter (baryonic and dark matter) ones.
Then, on a $4D$ hypersurface, we have classified the energy--momentum tensor into two parts.
One part represents the baryonic and dark matter together and the other one which contains every extra terms
emerging from the fifth dimension, have been considered as the energy--momentum tensor of dark energy.
Then, by considering a generalized FLRW metric in a $5D$ space--time, we have derived the FLRW cosmological
equations on the $4D$ hypersurface and separated the total energy conservation equation into the two equation,
one for the matter and the other one for dark energy with interacting term between them.
Afterwards, we have investigated three cases.

%%%%%%%%%%%%%%%%%%%%%%%%%%%%%%%%%%%%%%%%%%%%%%%%%%%%%%%%%%%%%%%%%%%%%
In the first one, we assumed the fifth--dimension momentum density be zero, $Q=0$.
In this case, for non--interacting situation for all kind of the geometry
the results are contrary to the observations and also for interacting one
the equations are not self consistent.
%%%%%%%%%%%%%%%%%%%%%%%%%%%%%%%%%%%%%%%%%%%%%%%%%%%%%%%%%%%%%%%%%%%%%
For the second case, we supposed the pressure in the fifth dimension equals to zero, i.e. $p_{_{5}}=0$.
The consistency condition between the equations yields to the same unsuitable results
of the first case with $Q=0$.
%%%%%%%%%%%%%%%%%%%%%%%%%%%%%%%%%%%%%%%%%%%%%%%%%%%%%%%%%%%%%%%%%%%%%

Finally, in the last case, we considered a tiny constant scale factor of the fifth dimension, i.e. $b=cte$.
In this situation by defining the effective energy densities of matter and dark energy we have illustrated
that in early time, the universe expands with deceleration and then in the late time, expands with acceleration.
The present acceleration, for this model, yields to $\Omega_{_{X\circ}}>2\Omega_{_{M\circ}}$ which is in agreement
with the measurements. The state parameter of the effective dark energy has the range $-1<\bar{w}_{_{X}}<0$ and
it is equal to $-1/2$ at the present time. We have also shown that the effective energy density of dark
energy have has the same order of magnitude of the effective energy density of matter since the early time until now.
Therefore, the model avoids the cosmological coincidence problem.

\section*{Acknowledgements}
\indent

I would like to thank Department of Physics and Astronomy, University of California,
Irvine, for the visit opportunity and their accommodations. Also, especial thanks to
Tim Tait and Arvind Rajaraman for reading this article and useful comments.

\makeatletter
\let\clear@thebibliography@page=\relax
\makeatother

\end{document}